\documentstyle[aps,prl,epsfig,twocolumn]{revtex}

\def\be{\begin{equation}}
\def\ee{\end{equation}}
\def\bea{\begin{eqnarray}}
\def\eea{\end{eqnarray}}
\def\bma{\begin{mathletters}}
\def\ema{\end{mathletters}}
\def\C{\hbox{$\mit /$\kern-.6em$\mit C$}}
\def\tr{{\rm tr}}
\newcommand{\eins}{\mbox{$1 \hspace{-1.0mm}  {\bf l}$}}

\begin{document}
\draft

\title{Distillability and partial transposition in bipartite systems}

\author{W. D\"ur$^1$, J. I. Cirac$^1$, M. Lewenstein$^2$ and D. Bru\ss$^2$}

\address{$^1$Institut f\"ur Theoretische Physik, Universit\"at Innsbruck,
A-6020 Innsbruck, Austria}
\address{$^2$ Institut f\"ur Theoretische Physik, Universit\"at Hannover,
D-30167 Hannover, Germany}

\date{\today}

\maketitle

\begin{abstract}
We study the distillability of a certain class of bipartite density operators 
which can be obtained via depolarization starting from an arbitrary one. Our 
results suggest that  non-positivity of the partial transpose of a density 
operator is not a sufficient condition for distillability, when the dimension of 
both subsystems is higher than two.
\end{abstract}


\narrowtext

\section{Introduction}

Maximally entangled states represent an essential ingredient in most 
applications of Quantum Information (QI)\cite{Ll95}. In particular, in Quantum 
Communication one can use them for transmitting secret messages between two 
locally separated parties \cite{Ek91}. In practice, however, states are mixed 
due to the interaction with the environment, and are not usable for those 
applications, even though they may be entangled. The solution to this problem 
was presented by Bennett {\it et al.}, Deutsch {\it et al.}, and Gisin 
\cite{Be96,De96,Gi96}, who have given  a procedure to ``distill'' maximally 
entangled states of two qubits out of a set of pairs in certain (mixed) 
entangled states, by only using local actions and classical communication 
\cite{Li98,Ke98}. Later on, the Horodecki family showed that any, even 
infinitesimally entangled state of two qubits (two-level systems) can be 
distilled into a singlet\cite{Ho97b}. They have also proved a necessary 
condition for the state of an arbitrary bipartite system to be distillable, 
namely, that the partial transpose of the corresponding density operator must be 
non-positive \cite{Ho98}. 

As shown by Peres \cite{Pe96}, the positivity of the partial transpose is a 
necessary condition for separability. In fact, this condition turns out to be a 
sufficient condition for separability  in both cases of qubits (two--level 
systems),  or one qubit and one trit (three--level system) \cite{Ho96}. A 
natural question arises: is this condition also sufficient for separability for 
higher dimensional systems? P. Horodecki has recently shown \cite{Ho97} that 
there are in fact states in higher dimensional systems which have a positive 
partial transpose, but are non separable (see also 
\cite{Ho98,Le98,Le99,Ho99,Be98,Li98b}). As a consequence, positivity of a partial 
transpose is, except for $2\times2$ and $2\times3$ systems, not sufficient for 
separability. Similarly, non-positivity of a partial transpose is necessary for 
distillability, and is sufficient for $2\times2$ and $2\times3$ systems. 
However,  the question whether this condition is sufficient for distillability 
in higher dimensional systems remains still open.

In this paper we investigate distillability of high dimensional systems shared 
by two parties, Alice and Bob. We introduce a depolarizing superoperator that 
allows one to reduce an arbitrary density operator with non positive partial 
transposition (NPPT) to one with the same property, but in a standard form that 
is characterized by a single parameter. We analyze some properties of those 
operators, and show that for any given finite number of copies there are density 
operators $\rho$ for which one can never find a subspace of dimension 2 in Alice 
and Bob's Hilbert spaces in which $\rho$ still has NPPT. We also present some 
numerical evidence that indicates that this class of states is independent of 
the number of copies. All these results suggest that there exist states with a 
NPPT and which are not distillable.

This work is organized as follows: In Section \ref{ED} we review
some of the present knowledge concerning distillability and
entanglement, and we introduce the definitions and properties
that are needed in order to study the problem of distillability
of general density operators. In Section \ref{Ndist3x3} we
concentrate on the case in which Alice and Bob have three--level
systems, whereas in Section \ref{Ndistdxd} we generalize our
results to the $d$--level system case. In Section \ref{NP} we
show the basics of the numerical procedure used to study the
distillability of 2 and 3 copies. Finally, we summarize our
results.

\section{Entanglement and distillability}
\label{ED}

We consider two parties, Alice and Bob, who share several pairs
of particles. Each pair is in a state described by the same
density operator $\rho$.  We will assume that
Alice's (Bob's) particles are $d_A$--level systems ($d_B$); that
is, the density operator $\rho$ acts on the Hilbert space
$\C^{d_A}\otimes\C^{d_B}$. We will denote by
$\{|1\rangle,|2\rangle,\ldots,|d_A\rangle\}$ an orthonormal
basis in $\C^{d_A}$ and analogously for $\C^{d_B}$. We will also
use the notation $|i,j\rangle\equiv
|i\rangle_A\otimes|j\rangle_B$.

We will assume that Alice and Bob are able to manipulate their
particles by only using local actions (operators and
measurements) and classical communication. In this case, we say
that the density operator $\rho$ is {\em distillable} if they
can produce a maximally entangled state
\be
\label{phid}
|\Phi_d\rangle = \frac{1}{\sqrt{d}}\sum_{i=1}^d |i,i\rangle,
\ee
where $d=\min(d_A,d_B)$. On the other hand, we say that $\rho$
is separable if it can be prepared out of a product state (e.g.
$|1,1\rangle$).

In this Section we will review some of the results derived by
Peres and the Horodecki family concerning distillability and
entanglement, and will introduce the definitions and properties
that are needed in order to study the problem of distillability
of general density operators.

\subsection{Partial transposition}
\label{pt}

As shown by Peres\cite{Pe96} and the Horodeckis \cite{Ho97b,Ho96,Ho97}, 
the partial 
transpose of a density operator plays an important role in establishing its 
distillability and entanglement properties. In general, given an operator $X$ 
acting on $\C^{d_A}\otimes\C^{d_B}$, we define the partial transpose of $X$ with 
respect to the first subsystem in the basis 
$\{|1\rangle,|2\rangle,\ldots,|d_A\rangle\}$, $X^{T_A}$, as follows:
\be
X^{T_A} \equiv \sum_{i,j=1}^{d_A}\sum_{k,l=1}^{d_B}
\langle i,k|X|j,l\rangle \; |j,k\rangle\langle i,l|.
\ee
In the following we will use a property of this operation,
namely $\tr(YX^{T_A})=\tr(Y^{T_A}X)$.

We say that a self--adjoint operator $X$ has a non--positive partial transpose 
(NPPT) if $X^{T_A}$ is not positive; that is, if there exist some 
$|\Psi\rangle$ such that $\langle\Psi|X^{T_A}|\Psi\rangle <0$. The positivity of 
the operator $\rho^{T_A}$ gives necessary criteria for separability and 
non--distillability of a density operator $\rho$. In particular: (1) If $\rho$ 
is separable, then $\rho^{T_A}\ge 0$ \cite{Pe96}; (2) If $\rho^{T_A}\ge 0$ then 
$\rho$ is not distillable \cite{Ho97}. These two necessary conditions turn out 
to be sufficient for $d_A=2$ and $d_B\le 3$\cite{Ho97b}. However, it has been 
shown that the first condition is not sufficient for separability for
the rest of the cases 
($d_A=2$ and $d_B>3$, and $d_A,d_B>2$)\cite{Ho97}. On the other hand, nothing is 
known about whether the second condition is also sufficient for non-distillability
 in these cases.

\subsection{Distillability}\label{dis}

The problem of distillability of general density operators
acting on $\C^{d_A}\otimes \C^{d_B}$ can be expressed in a
simpler form\cite{Ho99b}. A density operator is distillable iff for certain
positive integer $N$, we can find a state of the form
\be
\label{Psi}
|\Psi\rangle = a |e_1\rangle_A|f_1\rangle_B + b|e_2\rangle_A|f_2\rangle_B,
\ee
such that
\be
\label{cond}
\langle\Psi| (\rho^{\otimes N})^{T_A} |\Psi\rangle <0.
\ee
Here, $\{|e_1\rangle_A,|e_2\rangle_A\}$ are two orthonormal
vectors in $(\C^{d_A})^{\otimes N}$, and
$\{|f_1\rangle_B,|f_2\rangle_B\}$ are two orthonormal vectors in
$(\C^{d_B})^{\otimes N}$. This condition basically means that if
Alice and Bob share $N$ pairs, one just has to find a two-dimensional subspace in the whole Hilbert space of Alice, and
another in Bob's such that the projection of $\rho^{\otimes N}$
in such subspaces has NPPT. The reason is that if one finds such
a subspace, then according to what was exposed in the previous
subsection one can distill a maximally entangled state in
$\C^2\otimes\C^2$, which can be converted into a maximally
entangled state in $\C^{d_A}\otimes\C^{d_B}$. Conversely, if one
can create one of those states then one can also produce one in
$\C^2\otimes\C^2$, and therefore this ensures that (\ref{cond})
must be fulfilled.

Thus, in practice, one can analyze for each number of copies $N=1,2,\ldots$
whether the condition (\ref{cond}) is fulfilled. In order to
facilitate this task, we will use the following definitions: If
for a given $N$ condition (\ref{cond}) is fulfilled we will say
that $\rho$ is $N$--distillable. On the other hand, if for a
certain $N$ there {\it does not exis}t any  $|\Psi\rangle$ 
satisfying Eq.\
(\ref{cond}), then we will say that $\rho$ is $N$--undistillable.
Thus, $\rho$ is distillable iff there exists an  $N$ for which it is $N$-distillable.
Conversely, $\rho$ is non-distillable iff it is $N$--undistillable
$\forall N$.

\subsection{Distillability in $\C^2\otimes \C^d$}

With the properties and definitions given above, one can very
easily prove that when $d_A=2$ and $d_B\ge 2$ and if $\rho$ has
a NPPT then $\rho$ is 1--distillable. The reason is that there
exists some $|\Psi\rangle$ such that $\langle\Psi|
\rho^{T_A} |\Psi\rangle <0$. On the other hand, since $d_A=2$ then the Schmidt
decomposition of $|\Psi\rangle$ has at most two terms, and
therefore can be written in the form (\ref{Psi}). Thus, in this case 
non-positive partial transpose of $\rho$ is a necessary and
sufficient condition for distillability.

\subsection{Depolarization in $\C^d\otimes \C^d$}
\label{Depol}

In this subsection we will introduce some superoperators which
will be useful to study if a given density operator is
$N$--distillable. We will also show that given a density
operator one can reduce it to a standard form which is
characterized by a single parameter\cite{Ho99b}, and that preserves the
distillability properties of the original state.

Let us first define some useful projector operators. Given two
quantum systems with corresponding Hilbert spaces $\C^d$, we
denote by $\Pi_d$ the permutation operator, and by
\be
A_d=(\eins-\Pi_d)/2, \quad S_d=\eins-A_d=(\eins+\Pi_d)/2,
\ee
the projector operators onto the antisymmetric and symmetric
subspaces of $\C^d\otimes \C^d$, respectively. Note that ${\rm
tr}(S_d)=d(d+1)/2$ and ${\rm tr}(A_d)=d(d-1)/2$. We also define
the projector operators
\be
P_d=|\phi_d\rangle\langle\phi_d|,\quad Q_d=\eins-P_d,
\ee
where $|\phi_d\rangle$ is the maximally entangled state defined
in (\ref{phid}). One can easily check that
\be
P_d^{T_A}=\frac{1}{d} (\eins-2A_d), \quad A_d^{T_A} =\frac{1}{2} (\eins-
d P_d). \label{PdTa}
\ee

We define the depolarization superoperator ${\cal D}$, acting on
any given operator $X$, as follows:
\be
{\cal D}(X)= A_d \frac{{\rm tr}(A_d X)}{{\rm tr}(A_d)} + S_d
\frac{{\rm tr}(S_d X)}{{\rm tr}(S_d)}.\label{Dx}
\ee
This superoperator is a projector, is self--adjoint (on the
Hilbert--Schmidt space of operators acting on $\C^d\otimes
\C^d$), and preserves the trace. In the Appendix A we show that we
can write \cite{We89}
\be
\label{form}
{\cal D}(X) = \int d\mu_U (U\otimes U) X (U\otimes U)^\dagger
\ee
where the integral is extended to all unitary operators acting
on $\C^d$ and $\int d\mu_U = 1$ [$d\mu$ represents the standard
invariant Haar measure on the group $SU(d)$]. We will later on use the partial
transpose of ${\cal D}(X)$, and to this aim we define the
following superoperator
\begin{mathletters}
\bea
{\cal E}(X) &\equiv& [{\cal D}(X^{T_A})]^{T_A}
 = \int d\mu_U (U^\ast\otimes U) X (U^\ast\otimes U)^\dagger\\
&=& P_d {\rm tr}(P_d X) + Q_d \frac{{\rm tr}(Q_d X)}{{\rm
tr}(Q_d)}.
\eea
\end{mathletters}
where $U^\ast$ denotes complex conjugation in the basis in which
the partial transposition is defined. This superoperator is also a
projector, self--adjoint, and preserves the trace. Note that for
any unitary operator $V$ acting on $\C^d$ we have
\begin{mathletters}
\bea
(V\otimes V){\cal D}(X)(V\otimes V)^\dagger &=& {\cal D}(X),\\
\quad (V^\ast\otimes V){\cal E}(X)(V^\ast\otimes V)^\dagger &=&
{\cal E}(X).
\eea
\end{mathletters}

The form (\ref{form}) shows that the superoperators ${\cal D}$
and ${\cal E}$ can be implemented by means of local operations.
In particular, it shows that any density operator $\rho$ can be
transformed, using local operations, to the form
\be
\label{standard}
{\cal D}(\rho)=\rho_\alpha = \frac{1}{N(\alpha)} (S_d + \alpha
A_d)
\ee
where $\alpha$ is such that $\tr(A_d
\rho)=\tr(A_d\rho_\alpha)$ and $N(\alpha)={\rm tr}(S_d)+\alpha {\rm
tr}(A_d)$ is a normalization constant. That is, one can
depolarize any density operator to the one parameter family
(\ref{standard}) while keeping the weight in the antisymmetric
subspace. We will be more interested in the partial transpose of
$\rho_\alpha$, which is given by
\be
\label{rho_b}
[{\cal D}(\rho)]^{T_A} = {\cal E}(\rho^{T_A}) =
\frac{1}{M(\beta)} (Q_d - \beta P_d),
\ee
where $M(\beta)={\rm tr}(Q_d)-\beta$ is a normalization
constant, and the relationship between $\alpha$ and $\beta$ is
$\beta=[(\alpha-1)(d-1)-2]/(\alpha+1)<d-1$.
Note that since
\be
\rho_\alpha = {\cal D}(\rho_\alpha),\quad 
\rho_\alpha^{T_A} = {\cal E}(\rho_\alpha^{T_A}),
\ee
we have that
\begin{mathletters}
\bea
\label{prop2}
(U\otimes U)\rho_\alpha(U\otimes U)^\dagger &=&
\rho_\alpha,\\
\label{prop}
(U^\ast\otimes U)\rho_\alpha^{T_A}(U^\ast\otimes
U)^\dagger&=&\rho_\alpha^{T_A}.
\eea
\end{mathletters}
for any unitary operator $U$.

Using the properties derived above, one can easily check that
\be
\mbox{$\rho_\alpha$ is separable} \Leftrightarrow
\rho_\alpha^{T_A} \ge 0 \Leftrightarrow  \beta\le 0.
\ee
The last equivalence follows directly from (\ref{rho_b}). For
the first one we have: ($\Rightarrow$) See subsection
(\ref{pt}); ($\Leftarrow$) We have that for $\beta=0$ [i.e.
$\alpha=\alpha_0\equiv (d+1)/(d-1)= {\rm tr}(S_d)/{\rm
tr}(A_d)$]
\be
\rho_{\alpha_0}^{T_A} \propto Q_d \propto {\cal E}(|0,1\rangle\langle 0,1|),
\ee
which is obviously positive and separable, in which case
the same holds for $\rho_{\alpha_0}$. For $\beta<0$ ($\alpha<\alpha_0$) we
can always obtain $\rho_\alpha$ by adding the identity operator
(which is separable) to $\rho_{\alpha_0}$ (this is due to the
fact that ${\rm tr}(S_d)>{\rm tr}(A_d)$). 
\par Thus, for $\beta> 0$ ($\alpha>\alpha_0$) $\rho_\alpha$ is non--separable. One can easily
check that this condition is equivalent to
\be
{\rm tr}(A_d\rho_\alpha)> 1/2.
\ee

 This last form allows us to show that for any given density operator $\rho$ 
with NPPT, one can always transform it using local actions to the form 
(\ref{standard}) such that it still has NPPT\cite{Ho99b}. Let us show that. 
Suppose that for a given $|\Psi\rangle$, $\langle\Psi|\rho^{T_A}|\Psi\rangle 
<0$. We can write $|\Psi\rangle=\sum_{i=1}^{n\le d} c_i |u_i,v_i\rangle$ where 
${|u_i\rangle}_{i=1}^d$ and ${|v_i\rangle}_{i=1}^d$ form an orthonormal basis. 
The operator $\rho$ can be transformed by local operations to $\rho_s \propto 
(A^\dagger\otimes B^\dagger)\rho(A\otimes B)$, with ${\rm tr}(\rho_s^{T_A} 
P_n)<0$ (by simply taking $A=\sum_{i=1}^n |u_i^\ast\rangle\langle i|/c_i + 
\sum_{i=n+1}^d |u_i^\ast\rangle\langle i|$ and $B=\sum_{i=n+1}^d 
|v_i\rangle\langle i|$). Using that $0> {\rm tr}(\rho_s^{T_A} P_n)= {\rm 
tr}(\rho_s P_n^{T_A})$ and (\ref{PdTa}) we immediately obtain that ${\rm tr}(A_d 
\rho_s)\ge {\rm tr}(A_n \rho_s)\ge 1/2$. Since ${\cal D}$ conserves this 
quantity, we obtain that ${\cal D}(\rho_s)$ has a negative partial 
transpose.

As pointed out by the Horodecki \cite{Ho99b}, the problem of distillability can be reduced to the
study of density operators of the form (\ref{standard}). If we
find that all those operators with a NPPT are distillable, we
will have shown that NPPT is a necessary and sufficient
condition for distillability. On the contrary, if we find that
there exist an operator of the form (\ref{standard}) which has a
NPPT, but is not distillable, we will have shown that such a
condition is not sufficient. In the following Sections we will
show that there are density operators of the standard form with
a NPPT which are not $N$--distillable for certain values of $N$.
As we have seen in subsection \ref{dis} we can study that
by checking whether there exist vectors of the form (\ref{Psi})
fulfilling condition (\ref{cond}).

\section{$N$--distillability in $\C^3\otimes \C^3$}
\label{Ndist3x3}

We consider the case $d_A=d_B=3$ and a density operator of the
form
\be
\rho_\alpha = \frac{1}{N(\alpha)} (S + \alpha A),
\ee
where we have omitted the superindices $d=3$, and
$N(\alpha)=6+3\alpha$. According to the discussion in Subsection
(\ref{Depol}) we just have to consider $\alpha\ge\alpha_0 = 2$,
since otherwise $\rho_\alpha$ is separable. We also have for the
partial transpose
\be
\rho_\beta^{T_A} = \frac{1}{M(\beta)} (Q - \beta P)
\ee
where $\beta=(2\alpha-4)/(\alpha+1)$, with $2\ge\beta\ge 0$ and
$M(\beta)=8-\beta$.

\subsection{$1$--distillability}

We look for a vector of the form (\ref{Psi}) such that
(\ref{cond}) is fulfilled. Choosing $U$ such that
$U|e_{1,2}\rangle=|1,2\rangle$ and using the property
(\ref{prop}) we see that we can restrict ourselves to the
subspace spanned by $\{|1\rangle,|2\rangle\}$ in Alice's Hilbert
space. Defining by $\eins_{2}$ the projection operator into this subspace and by 
$\eins_{3}$ the identity operator in $\C^3$, we obtain after
projecting $\rho_\beta^{T_A}$ onto such subspace
\be
\eins_{2}^A \rho_\beta^{T_A}\eins_{2}^A \propto \eins_{2}^A\otimes \eins_{3}^B -
\frac{2(1+\beta)}{3} P_2
\ee
which is positive iff $\beta\le 1/2$. Thus, we obtain that
$\rho_\alpha$ is 1--distillable iff $\beta>1/2$ (or,
equivalently, $\alpha>3$).

\subsection{$2$--distillability}

Let us consider now two pairs, in a state $\rho_\alpha$. We will
show that for $\beta\le 1/4$  the state $\rho_\alpha$ is
$2$--undistillable. For any state $|\Psi\rangle$ of the form
(\ref{Psi}) we have
\bea
\langle \Psi|Q^{(1)}\otimes(Q^{(2)}-P^{(2)}/2)|\Psi\rangle= \nonumber \\
{\rm tr_2}[{\rm tr}_1(|\Psi\rangle\langle\Psi|Q^{(1)})
(Q^{(2)}-P^{(2)}/2)].
\eea
where the superscripts 1 and 2 refer to the first and second
pair, respectively. Using the fact that $Q$ is separable, and
therefore that it can be written as $Q=\sum_i
c_i|a_i,b_i\rangle\langle a_i,b_i|$ with $c_i>0$, we have
\be
{\rm tr}_1(|\Psi\rangle\langle\Psi|Q^{(1)})= \sum_i c_i
|\Psi_i\rangle\langle\Psi_i|
\ee
where $|\Psi_i\rangle = \langle a_i,b_i|\Psi\rangle$ is a
state acting on the second pair which itself has the form
(\ref{form}). Thus, according to the results of the previous
subsection we have that
$\langle\Psi_i|(Q^{(2)}-P^{(2)}/2)|\Psi_i\rangle
\ge 0$ and therefore $\langle
\Psi|Q^{(1)}\otimes(Q^{(2)}-P^{(2)}/2)|\Psi\rangle\ge 0$. In the same way we
have that $\langle \Psi|(Q^{(1)}-P^{(1)}/2)\otimes
Q^{(2)}|\Psi\rangle\ge 0$, i.e.
\be 0\le \langle \Psi|(Q^{(1)}-\frac{P^{(1)}}{4})\otimes(Q^{(2)}-\frac{P^{(2)}}{4})-P^{(1)}\otimes
\frac{P^{(2)}}{16}|\Psi\rangle.
\ee
Using the fact that $P^{(1)}\otimes P^{(2)}\ge 0$ we obtain the
desired result. Note that our results do not imply that for
$1/4\le\beta<1/2$ $\rho_\alpha$ is 2--distillable. In fact, as shown in
the next section, numerical calculations indicate that it is
2--undistillable.

\subsection{$N$--distillability}

We consider now $N$ pairs, in a state $\rho_{\alpha}$. We will show that for
$\beta \leq 4^{-N}$ the state $\rho_\alpha$ is $N$--undistillable. For any state
$|\Psi\rangle$ of the form (\ref{Psi}) one can check the following relations:
(i) $\langle\Psi|P^{\otimes k}|\Psi\rangle \leq \frac{2}{3^k}$,
(ii) $\langle\Psi|Q^{\otimes N-k}P^{\otimes k}|\Psi\rangle \leq \frac{2}{3^k}$ and
(iii) $\langle\Psi|Q^{\otimes N}|\Psi\rangle \geq \frac{1}{3^N}$. To show (i),
one uses $P_d^{\otimes k}=P_{d^k}$ and the property (\ref{prop}), from which
follows that the projection into the subspace spanned by
$\{|1\rangle,|2\rangle\}$ gives the maximum value for $\langle\Psi|P^{\otimes
k}|\Psi\rangle$. From (i) we immediately obtain (ii) by using that
$\langle\Psi|QX|\Psi\rangle \leq \langle\Psi|\eins X|\Psi\rangle$ for all
positive operators $X$. Relation (iii) can be obtained by using (i) and the
separability of Q in a similar way as in the previous Section. Combining (ii)
and (iii) we find \be \langle\Psi|a_k Q^{\otimes N} -\sum_{\rm perm} Q^{\otimes
N-k} P^{\otimes k}|\Psi\rangle \geq 0,\label{el4} \ee with $a_k=2 {N \choose k}
3^{N-k}$ and the sum runs over all possible permutations of the pairs. By
summing (\ref{el4}) for all odd $k$ and using that $\sum_{k \, \rm odd}a_k=4^N-2^N
\equiv \tilde\beta_N^{-1}$, one finds for $\beta_N \leq \tilde\beta_N$
\bea
0 &\leq& \langle\Psi|Q^{\otimes N} -\tilde\beta_N\sum_{k \, \rm
odd}\sum_{\rm perm} Q^{\otimes N-k} P^{\otimes k}|\Psi\rangle
\nonumber\\ &\leq&\langle\Psi|Q^{\otimes N} -\sum_{k \, \rm
odd}\beta_N^k
\sum_{\rm perm} Q^{\otimes N-k} P^{\otimes k}|\Psi\rangle \nonumber \\
&\leq&\langle\Psi|(Q-\beta_N P)^{\otimes N}|\Psi\rangle. \label{inequ}
\eea
We used that ${\rm max}_k (\beta_N^k) \leq \tilde\beta_N$  (for
$\beta_N\leq\tilde\beta_N\leq 1$) to obtain the first inequality (line 2), while we added
all positive terms (even $k$) in the second step (line 3). This is already the
desired bound, i.e. for $0\leq \beta\leq 4^{-N}\leq \tilde\beta_N\equiv
\frac{1}{4^N-2^N}$, the state $\rho_\alpha$ is $N$--undistillable. 
Again, this does not mean that $\rho_\alpha$ is $N$--distillable for $4^{-N} \leq \beta \leq 1/2$.

In the Appendix C we present a better bound for $\beta$.

\section{$N$--distillability in $\C^d\otimes\C^d$}
\label{Ndistdxd}

We consider the case $d_A=d_B=d$ and a density operator of the form
(\ref{standard}) with the partial transposition given by (\ref{rho_b}).

Similar techniques as in the $d=3$ case can be used to obtain bounds also for
arbitrary $d$. One finds for example that $\rho_\alpha$ is 1--distillable iff
$\beta > \frac{d}{2}-1$. We also obtain that $\rho_\alpha$ is 2--undistillable
for $\beta \leq \frac{d-2}{4}$ and $N$-undistillable for $\beta \leq {\rm
min}(\tilde\beta_N,\tilde\beta_N^{1/N})$ with
$\tilde\beta_N=\frac{(d-2)^N}{(d+1)^N-(d-1)^N}$. Note that the
minimum is required here, since - differently to $d$=3 case - one can have that
$\tilde\beta_N \geq 1$. In this case one has to chose $\beta_N \leq
\tilde\beta^{1/N}$ which implies ${\rm max}_k\beta_N^k \leq \tilde\beta_N$ to ensure
that the first inequality in (\ref{inequ}) remains valid.

Furthermore, there is an interesting relation between  the states
$\rho_{\alpha}$ for different $d$. Imagine we would like to convert a single
copy of a state $\rho_\alpha$ in $\C^{d}\otimes\C^{d}$ to a state in
$\C^k\otimes\C^k$ ($k < d$) in an optimal way, i.e. to obtain a new $\alpha$
which is as large as possible. We show here that whenever we convert a state
$\rho_\alpha$ to some lower dimension, there will always be some states which loose the
negativity of their partial transposition. In order to prove this, we consider
vectors $|\Psi_k\rangle$ with $k$ Schmidt coefficients and show that
$\langle\Psi_d|\rho_\alpha^{T_A}|\Psi_d\rangle < 0$, while
$\langle\Psi_k|\rho_\alpha^{T_A}|\Psi_k\rangle > 0 \, \forall \, |\Psi_k\rangle$. Due
to the property (\ref{prop}), one can restrict oneself to the subspace spanned by
$\{|1\rangle \ldots |k\rangle\}$. Let us denote the identity operator in this
subspace by $\eins_k$. One finds after projecting $\rho_\alpha^{T_A}$ onto $\eins_k$ in
$A$ and $B$
\be
\eins_{k}^A\otimes \eins_{k}^B \rho_\alpha^{T_A} \eins_{k}^A\otimes \eins_{k}^B \propto
\eins_{k}^A\otimes \eins_{k}^B-\frac{k(1+\beta)}{d}P_k
\ee
which is positive iff $\beta \leq \frac{d}{k}-1$, while $\rho_\alpha^{T_A}$ 
before the projection was positive iff $\beta \leq 0$. Thus all states with $0 < 
\beta \leq \frac{d}{k}-1$ lose the negativity of their partial transposition 
after the optimal projection onto a $k$-dimensional subspace. The new $\beta_k$ 
can be calculated from $\beta_d$ of the initial state by 
$\beta_k=\frac{k}{d}(\beta_d+1)-1$. 

Finally, let us consider $N$ copies of $\rho_\alpha$ of dimension $d$, which can 
be viewed as a state in $\C^{\otimes d^N}\otimes\C^{\otimes d^N}$. With ${\rm 
tr}(A_d\rho_\alpha) \equiv \lambda_d$, one finds that the state ${\cal 
D}(\rho_\alpha^{\otimes N}) \equiv \tilde\rho_\alpha$ - the state in the high 
dimensional Hilbert space after depolarization - has $\lambda_{d^N}\equiv {\rm 
tr}(A_{d^N} \tilde\rho_\alpha) = \frac{1-(1-2\lambda_d)^N}{2}$.                          
One checks that for $\lambda_d>1/2$ (i.e. $\rho_\alpha$ is inseparable) we have 
$\lambda_{d^N} \leq \lambda_d \forall N$, which simply means that the weight in the 
antisymmetric subspace decreases when going to more copies. Note that using this 
notation, we have that the state $\rho_\alpha$ is separable for $\lambda \leq 1/2$, 
while it is 1-distillable for $\lambda>\frac{3(d-1)}{2(2d-1)}$, which tends to 
$\frac{3}{4}$ for $d \rightarrow \infty$.

\section{Numerical Procedures}
\label{NP}

In general, we are interested in showing there exists a
$|\Psi\rangle$ of the form (\ref{Psi}) for which condition
(\ref{cond}) is fulfilled, i.e.
\be
\lambda\equiv \langle \Psi|R|\Psi\rangle <0,
\ee
where we have defined $R=(Q-\beta P)^{\otimes N}$ with
$0<\beta\le 1/2$. In order to check that, we can minimize
$\lambda$ with respect to $|e_{1,2}\rangle$, $|f_{1,2}\rangle$
and $a$ while keeping the normalization and orthogonality
relations. One can readily check that the minimization implies
\begin{mathletters}
\bea
\label{M1}
\langle e_1|R|\Psi\rangle &=& \lambda_0 a|f_1\rangle,\\
\label{M2}
\langle e_2|R|\Psi\rangle &=& \lambda_0 b|f_2\rangle,\\
\langle f_1|R|\Psi\rangle &=& \lambda_0 a|e_1\rangle,\\
\langle f_2|R|\Psi\rangle &=& \lambda_0 b|e_2\rangle.
\eea
\end{mathletters}

Note that the operator $_A\langle e_1|R|e_1\rangle_A$ is
strictly positive, and therefore invertible. The reason for that
is that for any $|f\rangle_B$, we have $\langle e,f|R|e,f\rangle
=\langle e^\ast,f|R^{T_A}|e^\ast,f\rangle>0$ since according to
(\ref{standard}) we can always write $R^{T_A}=c\eins + B$ (where $B\ge
0$ and $c>0$). On the other hand, $a,b\ne 0$ since
otherwise $|\Psi\rangle$ would be a product vector and therefore
$\lambda\ge 0$. Thus, we can use (\ref{M1}) to write
\be
|f_1\rangle = \frac{b}{a} \frac{1}{\lambda_0-\langle
e_1|R|e_1\rangle}\langle e_1|R|e_2\rangle \; |f_2\rangle 
\ee
which, after substituting in Eq. (\ref{M2}) gives
\be
\label{baseq}
\langle e_2|F(\lambda_0)|e_2\rangle \; |f_2\rangle = \lambda_0
|f_2\rangle,
\ee
where we have defined
\be
F(\lambda_0) = R - R|e_1\rangle\frac{1}{\langle e_1|R|e_1\rangle
- \lambda_0}\langle e_1|R.
\ee
The normalization of $|f_1\rangle$ gives
\be
\left|\frac{a}{b}\right|^2 = \langle e_2,f_2| \left[
\frac{1}{\langle e_1|R|e_1\rangle - \lambda_0} \right]^2
|e_2,f_2\rangle.
\ee

Thus, the problem is reduced to showing whether Eq.\
(\ref{baseq}) possesses solutions for $\lambda_0<0$. In that
case, we can find $a,b,|f_1\rangle$ using the other
equations. On the other hand, if we have that
\be
\langle e_2|F(0)|e_2\rangle \ge 0
\ee
for all $|e_2\rangle$, then we will have that there exists no
solution with $\lambda_0<0$. This is so since
$F(-|\lambda_0|)-F(0)\ge 0$.

We have made a systematic search of the states $|e_{1,2}\rangle$ which minimize 
the minimum eigenvalue of $\langle e_2|F(0)|e_2\rangle \ge 0$ for $d=3$ and $N=2,3$. Note 
that for $N=2$ copies we can simplify further the
numerical search by using the symmetries of the problem, which
imply that we can choose $|e_1\rangle=\sum_{i=1}^3
c_i|i,i\rangle$ with $c_i \geq 0$.

In 
both cases we have found that this minimum eigenvalue is $\ge 0$ for $\beta\le 
1/2$, which strongly indicates that $\rho_\alpha$ is 3--undistillable (and hence also 
2--undistillable) for $\beta \leq 1/2$. This is exactly the same bound that  
we had obtained analytically  for 1--undistillability.

\section{Conclusions}

\begin{figure}[ht]
\begin{picture}(230,110)
\put(0,5){\epsfxsize=230pt\epsffile[42 699 265 801]{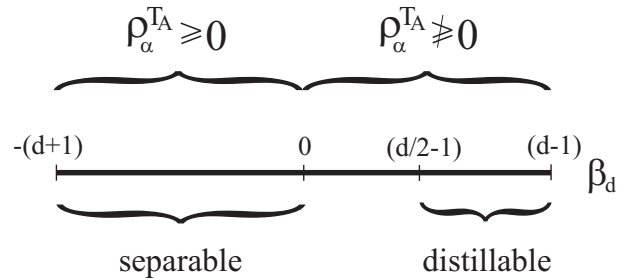}}
\end{picture}
\caption[]{Separability and distillability properties of $\rho_\alpha$ (\ref{standard}).}
\label{Fig1}
\end{figure}

We have shown that in order to study the distillability properties of bipartite 
$d$--level systems, it is sufficient to consider only the one--parameter class 
of states $\rho_\alpha$ (\ref{standard}). By investigating the distillability 
properties of this family of states, we found strong indications that this 
family provides examples for non--distillable states with non-positive partial 
transposition. In particular, we found that for any given number of copies $N$ there exist 
$N$--undistillable states which have NPPT. Guided by the results of the numerical 
investigations, we conjecture that for $d=3$ the states $\rho_\alpha$ are 
non--distillable for $\beta \leq 1/2$, while they have NPPT for $\beta > 0$
 (see also Fig.\ref{Fig1}).
\\
\\
Note added: After completing this work we became aware of the results of D. P. 
DiVincenzo {\it et.al} \cite{Di99}, in which they also found evidences for the 
existence of non--distillable states with NPPT.


\section{acknowledgments}
W. D. thanks the University of Hannover for hospitality.
This work was supported by the Deutsche Forschungsgemeinschaft under SFB
407, \"Osterreichischer Fonds zur F\"{o}rderung
der wissenschaftlichen Forschung, the European Community under the TMR
network ERB--FMRX--CT96--0087, the 
European Science Foundation and the
Institute for Quantum Information GmbH.


\section*{Appendix A: Integral representation of ${\cal D}(X)$}

We show here that the superoperator ${\cal D}$ as defined in (\ref{Dx}) can also 
be written in the form (\ref{form}). We restrict ourselves here to operators $X$ 
which are density operators $\rho$ for convenience, but exactly the same line of 
arguments holds for arbitrary self adjoint-operators $X$. As shown in 
Appendix B, we have that the depolarization superoperator ${\cal D}$ can be 
implemented by a finite sequence of bi--local operations (\ref{dep}) of the form 
$U\otimes U$. Furthermore we have that the projector onto the antisymmetric 
subspace is invariant under unitary operations of the form $U\otimes U$, i.e 
$U\otimes U A_d U^\dagger\otimes U^\dagger = A_d$, which can be easily seen by 
using that $A_d=(1-\Pi_d)/2$. From this property automatically follows that also 
$\rho_\alpha$ (\ref{standard}) is invariant under unitary 
operations of the form $U\otimes U$, 
since $\rho_\alpha \propto (\eins+\tilde\alpha A_d)$. It is now straightforward to 
show (\ref{form}) by using for any $V$ 
\bea
&&\int d\mu_U (U\otimes U) \rho (U\otimes U)^\dagger =  \nonumber \\
&&\int d\mu_{U'} (U'\otimes U') (V\otimes V) \rho (V\otimes V)^\dagger (U'\otimes U')^\dagger. \label{UU}
\eea
with $U'V=U$.  Taking $p_k$ such that $\sum_k p_k =1$ we can write
\bea
&&\int d\mu_U (U\otimes U) \rho (U\otimes U)^\dagger = \nonumber \\
&&\sum_k p_k \int d\mu_{U'} (U'\otimes U') (U_k\otimes U_k) \rho (U_k\otimes U_k)^\dagger (U'\otimes U')^\dagger = \nonumber \\
&&\int d\mu_{U'} (U'\otimes U') \rho_\alpha (U'\otimes U')^\dagger = \rho_\alpha = {\cal D}(\rho),
\eea
where we used (\ref{UU}) in the first equality, while the second equality 
follows from (\ref{dep}) and we finally used the invariance of $\rho_\alpha$ 
under operations of the form $U\otimes U$. This already shows that (\ref{form}) 
is fulfilled, i.e. ${\cal D}(\rho) = \int d\mu_U (U\otimes U) \rho (U\otimes 
U)^\dagger$.


\section*{Appendix B: Depolarization}
We are going to show now that an arbitrary state $\rho$ can be depolarized to  the 
standard form (\ref{standard}) by a finite sequence of bi--local (random) operations 
without changing the weight in the antisymmetric subspace ${\cal H}_a(d)$, i.e we show 
that there exist unitary operators $U_k$ and probabilities $p_k$ such that 
\be
\sum_k p_k U_k\otimes U_k \rho U_k^\dagger\otimes U_k^\dagger = {\cal D}(\rho) =\rho_\alpha \label{dep}
\ee
with ${\rm tr}(A_d \rho) = {\rm tr}(A_d \rho_\alpha)$.
We first introduce a basis of $\C^d\otimes\C^d$: 
\bea
|\varphi_{ij}^{\pm}\rangle&=&\frac{1}{\sqrt{2}}(|i\rangle_A|j\rangle_B \pm |j\rangle_A|i\rangle_B) \nonumber \\
|\chi_{k}\rangle&=&|k\rangle_A|k\rangle_B, \label{basis}
\eea
with $i<j$ and $(i,j,k) \in \{1,...,d\}$. Note that $|\varphi_{ij}^{-}\rangle$ provides 
a basis of the antisymmetric subspace ${\cal H}_a(d)$, while $\{|\varphi_{ij}^{+}\rangle,|\chi_{k}\rangle\}$
is a basis of the symmetric subspace ${\cal H}_s(d)$. The projectors 
into the symmetric/antisymmetric subspace can thus be written as
\bea 
A_d&=&\sum_{i,j=1 (i<j)}^{d}|\varphi_{ij}^{-}\rangle\langle
\varphi_{ij}^{-}| \nonumber \\ 
S_d&=&\sum_{i,j=1 (i<j)}^{d} |\varphi_{ij}^{+}\rangle\langle \varphi_{ij}^{+}| + 
\sum_{k=1}^{d}
|\chi_{k}\rangle\langle \chi_{k}|. 
\eea 
Let us write $\rho$ in the basis 
(\ref{basis}). In order to prove the statement (\ref{dep}), we devide the 
depolarization procedure into three steps: 
\begin{itemize}
\item we show that one can make $\rho$ diagonal in the basis (\ref{basis}) 
without changing the diagonal elements. 
\item we prove that the antisymmetric subspace ${\cal H}_a(d)$ can be mixed up, i.e. 
one can equalize the coefficients of   
$|\varphi_{ij}^-\rangle\langle\varphi_{ij}^-|$ without changing the weight in 
${\cal H}_a(d)$.
\item finally we show that also the symmetric subspace can be completely mixed 
up without changing the weight in ${\cal H}_s(d)$.
\end{itemize}
These three steps together ensure that $\rho$ can be depolarized to  the 
standard form $\rho_\alpha$.

\subsection{Diagonalizing $\rho$}
By mixing we understand in the following that a certain operation U is (randomly) performed 
with probability $p$ by Alice and Bob, while with probability $1-p$ no operation 
is performed. The resulting density operator 
after this mixing--operation reads
\be
\rho_{\text{new}}=p( U \otimes U \rho U^{\dagger} \otimes U^{\dagger}) + (1-p)\rho.
\ee
We define the operation $U_l$ as follows: 
\be 
U_l|k\rangle = e^{i\pi \delta_{kl}}|k\rangle,
\ee
i.e. the state $|l\rangle$ picks up a minus sign while all others remain unchanged. 
Let us perform a sequence of 
$d$ mixing operations, using p=$\frac{1}{2}$ and $U_l$ with $l= \{1,...,d\}$ respectively. One 
can easiliy check that all diagonal elements remain unchanged, while all 
off--diagonal elements of the form $|\varphi_{ij}^\pm\rangle\langle\chi_k|$ and 
$|\varphi_{ij}^\pm\rangle\langle\varphi_{i'j'}^\pm|$ for $(i \neq i',j \neq j')$ 
are eliminated. 

Let us define now the operation $\hat U_l$, which introduces a phase $i$ for the 
state $|l\rangle$ while it leaves all other states unchanged. Performing again 
the same sequence of $d$ mixing operations as described above, but with $U_l$ 
replaced by $\hat U_l$, one can check that all off--diagonal elements of the 
form $|\chi_i\rangle\langle \chi_j|$ are eliminated.

We finally define the operation $U_{kl}$, which simply swaps the states 
$|k\rangle$ and $|l\rangle$ while leaving all other states untouched. Performing 
now a sequence of mixing operations using all possible combinations $k < l \in 
\{1,...,d\}$ for $U_{kl}$ and $P=1/2$, one gets rid of the remaining off--diagonal 
elements of the form $|\varphi_{ij}^-\rangle\langle\varphi_{ij}^+|$ without 
changing the diagonal ones. Thus $\rho$ can be made diagonal in the basis 
(\ref{basis}) by a sequence of bi--local operations.


\subsection{Mixing of ${\cal H}_a(d)$}
Let ${\cal H}_{i_0}$ be the subspace spanned by $\{|\varphi_{i_0j}^-\rangle\}$ for 
fixed $i_0$. In a first step we will show that one can depolarize all subspaces 
${\cal H}_{i_0}$ independently, while in the second step we are going to prove that 
these subspaces can be mixed with each other.

To depolarize ${\cal H}_1$, one just has to keep the state $|1\rangle$ and performs 
(randomly) one of the cyclic permutations of the states 
$\{|2\rangle,...,|d\rangle\}$ each with probability $p=\frac{1}{d-1}$. 
Similary, one depolarizes ${\cal H}_2$ by keeping the states $\{|1\rangle,|2\rangle\}$ 
and performing with probability $p=\frac{1}{d-2}$ one of the cyclic permutations 
of the states $\{|3\rangle,...,|d\rangle\}$. Since ${\cal H}_1$ is already 
depolarized, it is not affected by this operation. One can continue in the same 
way until one has depolarized ${\cal H}_{d-1}$.  So the antisymmetric part of the 
density operator has now the form 
\be
A_d \rho A_d=\sum_{i=1}^{d-1} a_i \sum_{j=i+1}^{d}
|\varphi_{ij}^{-}\rangle\langle\varphi_{ij}^{-}|.  
\ee

The second step starts by mixing of ${\cal H}_{d-1}$ with ${\cal H}_{d-2}$, i.e. 
equalizing the coefficients $a_{d-1}$ and $a_{d-2}$. To achieve this, both Alice 
and Bob swap the states $|d-1\rangle$ and $|d-2\rangle$ with probability 
$p_s=\frac{2}{3}$, or both apply the identity operator with probability  $1-p_s$. 
If one now depolarizes ${\cal H}_{d-1}$ and ${\cal H}_{d-2}$ independently as 
described in step 1, one finds that 
$|\varphi_{(d-1)j}^-\rangle\langle\varphi_{(d-1)j}^-|$ and 
$|\varphi_{(d-2)j}^-\rangle\langle\varphi_{(d-2)j}^-|$ have all the same weight 
now, i.e. the coefficients are equal. Thus the subspace  ${\cal H}_{d-1}$ is 
completely mixed up with ${\cal H}_{d-2}$. One now continues by mixing the 
subspaces ${\cal H}_{d-3}$ with $\{{\cal H}_{d-2},{\cal H}_{d-1}\}$ and so on, 
until one reaches ${\cal H}_1$.

We investigate now one particluar step in this procedure, namely the mixing of ${\cal H}_{k}$ with 
$\{{\cal H}_{k+1},{\cal H}_{k+2},....,{\cal H}_{d-1}\}$. Both Alice and Bob swap the states  
$|k\rangle$ and $|k+1\rangle$ with probability $p_s=\frac{d-k}{d-k+1}$ or both 
apply the identity operator with probability  $1-p_s$.  After this one 
depolarizes ${\cal H}_j  (j \leq k)$ independently, then mixes ${\cal H}_{d-1}$ with  
${\cal H}_{d-2}$ as described above. Next one mixes  ${\cal H}_{d-3}$ with 
$\{{\cal H}_{d-2},{\cal H}_{d-1}\}$ and continues in this way until one has mixed ${\cal H}_k$ and 
$\{{\cal H}_{k+1},{\cal H}_{k+2},....,{\cal H}_{d-1}\}$ . It can be checked that after this procedure 
all weight factors $\tilde{a_j}$ are equal for ($j \leq k$).

So once one has mixed ${\cal H}_1$ with $\{{\cal H}_{2},{\cal H}_{3},....,{\cal H}_{d-1}\}$, one has 
achieved that the whole antisymmetric subspace is completely depolarized, i.e. 
it can be written as $\alpha_d A_d$.

\subsection{Mixing of ${\cal H}_s(d)$}
Note that the depolarization of the antisymmetric subspace also mixes the subspace 
spanned by $\{|\varphi_{ij}^+\rangle\}$ in a similar way. Here we show now that 
one can also depolarize the subspace spanned by  $\{|\chi_{k}\}$ and finally 
that also these two subspaces together can be mixed up. To prove the first step, 
let us define the operation $\tilde U_l$ as follows: 
\be
\tilde U_l|k>=|(k+l)\text{mod}(d)>.
\ee
Performing now $\tilde U_l \otimes \tilde U_l, l=\{1,....,d\}$ each with 
probability p=$\frac{1}{d}$ ensures that the subspace spanned by 
$\{|\chi_{k}\rangle\}$ is completely depolarized, i.e. that $\rho$ has now the 
form
\be
\rho=a A_d + b  \sum_{i,j=1 (i<j)}^{d} |\varphi_{ij}^{+}\rangle\langle 
\varphi_{ij}^{+}| + c \sum_{k=1}^{d} |\chi_{k}\rangle\langle \chi_{k}|.
\ee
For the second step, we define the unitary operation $T$:
\be 
T|j\rangle = \frac{1}{\sqrt{d}}\sum_{k=1}^{d} e^{i2\pi (j-1)(k-1)/d} |k\rangle.
\ee

One can check that if we perform the operation $T \otimes T$ with probability 
$p_T=\frac{d}{d+1}$ and the identity operation with probability $p_I=1-p_T$, the 
diagonal elements of the symmetric subspace will each be identical to 
$\frac{bd(d-1)+2cd}{d(d+1)}$. The introduced off--diagonal elements can be 
eliminated using the procudure explained above. Note that the antisymmetric 
subspace is not affected by this kind of operations and will thus remain 
untouched. So finally we managed to show that $\rho$ can be converted to the 
standard form (\ref{standard}) by a sequence of local operations (\ref{dep}). 
The weight in ${\cal H}_a(d)$ was not affected by any of the used mixing 
operations, which ensures that ${\rm 
tr}(A_d \rho) = {\rm tr}(A_d \rho_\alpha)$.

\section*{Appendix C: Better bound for  $\beta$}
In this Appendix we prove that for any vector  $|\Psi\rangle$ of the 
form (\ref{Psi}) 
\be
\langle \Psi|(Q-\beta P)^{\otimes N}|\Psi\rangle 
\ge \frac{1}{3^N}\left(1-\frac{\beta}{\beta_N}\right),
\ee
for $\beta\le\beta_N$, where 
\be
\beta_N=\frac{x_N}{3^{N/3}N^{1/3}},
\ee
and $x_N=x^*(1-O(1/N))$ as $N\to \infty$, where
$x^*=3(1-3^{-1/3})^{1/3}$, and $O(1/N)$ denotes a quantity of the order of
$1/N$.

The proof is by induction. For $N=1$, we have $\beta_1=1/2$. Assuming that
the statement holds for $N-1$, we observe that
\be
\langle \Psi|Q(Q-\beta P)^{\otimes N-1}|\Psi\rangle \ge
\frac{1}{3^N}\left(1-\frac{\beta}{\beta_{N-1}}\right),
\ee
and the same holds 
for all possible permutations with respect to the copies.
Adding LHS's, dividing the sum  by $N$, and rearranging various terms, 
we find for large $N$
\begin{eqnarray}
&\langle& \Psi|(Q-\frac{N-1}{N}\beta P)^{\otimes N}|\Psi\rangle \ge\\ 
&&\frac{1}{3^N}\left(1-\frac{\beta}{\beta_{N-1}}\right) - \frac{N\beta^3}{27}\left(1+ O(1/N)\right)
-O(\beta^5).\nonumber
\end{eqnarray}
For $\beta\simeq O(\beta_N)$, the last correction term in the above inequality
can be safely neglected.
It is easy to check that the right hand side has zero at 
$\beta=\beta_N$, i.e. that
\be
\langle \Psi|(Q-\beta P)^{\otimes N}|\Psi\rangle \ge \frac{1}{3^N}\left(1-
\frac{\beta}{\beta_N}\right),
\ee
for $\beta\le\beta_N$; the statement holds thus for every $N$. 

The above  result provides a better bound for $N$-undistillability:
the states $\rho_{\alpha}$ such that 
\be
\beta\le 
\frac{x^*}{3^{N/3}N^{1/3}}(1-O(1/N))
\ee
 are $N$-undistillable.



\begin{references}
\bibitem{Ll95}
S. Lloyd, Scientific American Oct. 1995, p. 140; 
\newline C. H. Bennett, Phys. Today, Vol. 24 (Oct. 1995).

\bibitem{Ek91}
A. K. Ekert, Phys.\ Rev.\ Lett. {\bf 70}, 661-663 (1991).


\bibitem{Be96}
C. H. Bennett, G. Brassard, S. Popescu, B. Schumacher, J. A. Smolin and W. K. Wootters, Phys. Rev. Lett. {\bf 76}, 722 (1996);
\newline C. H.Bennett, D. P. DiVincenzo, J. A. Smolin and W. K. Wootters, Phys. Rev. A {\bf 54}, 3824 (1996).

\bibitem{De96}
D. Deutsch, A. Ekert, C. Macchiavello, S. Popescu, and A. Sanpera,
Phys. Rev. Lett. {\bf 77}, 2818 (1996).

\bibitem{Gi96}
N. Gisin, Phys. Lett. A {\bf 210} 151 (1996).


\bibitem{Li98}
N. Linden, S. Massar and S. Popescu, Phys. Rev. Lett. {\bf 81}, 3279 (1998).

\bibitem{Ke98} A. Kent, Phys. Rev. Lett. {\bf 81}, 2839 (1998).


\bibitem{Ho97b}
M. Horodecki, P. Horodecki and R. Horodecki, Phys. Rev. Lett. {\bf
78}, 574 (1997).

\bibitem{Ho98}
M. Horodecki, P. Horodecki and R. Horodecki, Phys. Rev. Lett. {\bf 80}, 
5239 (1998).


\bibitem{Pe96}
A. Peres, Phys. Rev. Lett. {\bf 77}, 1413 (1996).

\bibitem{Ho96} M. Horodecki, P. Horodecki and
R. Horodecki, Phys. Lett. A{\bf 223}, 8 (1996).


\bibitem{Ho97}
P. Horodecki, Phys. Lett. A {\bf 232}, 333 (1997).
 
 \bibitem{Le98}
M. Lewenstein and A. Sanpera, Phys. Rev. Lett. {\bf 80}, 2261 (1998).

\bibitem{Le99}
M. Lewenstein, J. I. Cirac and S. Karnas, quant-ph/9903012.

\bibitem{Ho99}
M. Horodecki, P. Horodecki and R. Horodecki, Phys. Rev. Lett. {\bf 82},
1056 (1999).

\bibitem{Be98} 
C. H. Bennett, D. P. DiVincenzo, TalMor, P. W. Shor, J. A. Smolin and B. Terhal, Phys. Rev. Lett. {\bf 82}, 5385 (1999).

\bibitem{Li98b}
N. Linden and S. Popescu, quant-ph/9807069. 

\bibitem{Ho99b}
M. Horodecki and P. Horodecki, Phys. Rev. A. {\bf 59},
4206  (1999).

\bibitem{We89}
R. F. Werner, Phys. Rev. A {\bf 40}, 4277 (1989).

\bibitem{Di99}
D. P. DiVincenzo, R. Jozsa, P. W. Shor, J. A. Smolin, B. M. Terhal and A. V. Thapliyal, in preparation

\end{references}
\end{document}